\documentclass[a4paper,11pt]{article}
\pdfoutput=1 

\usepackage{jcappub} 

\usepackage[T1]{fontenc} 

\title{\boldmath Imprints of PeV cosmic-ray sources on the diffuse gamma-ray emission}

\author[a,b,1]{Samy Kaci\note{Corresponding author.}}
\author[a,b,c,1]{and Gwenael Giacinti}

\affiliation[a]{Tsung-Dao Lee Institute, Shanghai Jiao Tong University,\\ Shanghai 201210, P. R. China}
\affiliation[b]{School of Physics and Astronomy, Shanghai Jiao Tong University,\\ Shanghai 200240, P. R. China}
\affiliation[c]{Key Laboratory for Particle Physics, Astrophysics and Cosmology (Ministry of Education) \& Shanghai Key Laboratory for Particle Physics and Cosmology\\ 800 Dongchuan Road, Shanghai, 200240, P. R. China}

\emailAdd{samykaci@sjtu.edu.cn}
\emailAdd{gwenael.giacinti@sjtu.edu.cn}

\abstract{We present our new model for the description of the very high energy Galactic gamma-ray emission based on a discrete injection of cosmic rays by individual sources. We investigate the morphology of the very high energy gamma-ray sky, the detectability of cosmic-ray sources and the clumpiness of the diffuse gamma-ray flux, assuming two different scenarios for cosmic-ray propagation. Namely, a standard isotropic and homogeneous diffusion process and an isotropic and inhomogeneous diffusion process. We notably formulate a possible explanation to the small number of hadronic PeVatrons recently detected by LHAASO. In the case of the inhomogeneous diffusion process, we constrain the number of hadronic PeVatrons to be small. Finally, we give an argument that may explain the discrepancy between the interstellar gas density distribution and the very high energy diffuse gamma-ray flux.}

\begin{document}
\maketitle
\flushbottom
\section{Introduction}

The recent advent of the new generation of gamma-ray observatories opened a new observational window on our Galaxy, providing substantial amounts of very high energy (VHE) gamma-ray data. The AS$\gamma$ collaboration detected its first sub-$\rm{PeV}$ photons and published its first measurement of the VHE diffuse gamma-ray flux \cite{tibet1, tibet2}, HAWC reported several $\rm{TeV}$ cosmic-ray sources out of which three still emit at energies higher than $100\,\rm{TeV}$ \cite{hawc}, and the LHAASO collaboration published its data on the VHE diffuse gamma-ray background \cite{lhaaso1} and its first source catalog \cite{catalog}. In addition, IceCube provided the neutrino counterpart of the VHE Galactic gamma-ray observations \cite{icecube}, allowing for multi-messenger studies of our Galaxy. In this context, a picture of the VHE gamma-ray sky starts to emerge and raises new questions. While current models allow for a satisfying description of the Galaxy at relatively low energies ($\rm{GeV}$ to $\rm{TeV}$), addressing questions such as determining the nature of the so-called PeVatrons or the origin of the VHE diffuse gamma-ray emission requires more sophisticated approaches.

In this regard, we present a description of the VHE gamma-ray emission of our Galaxy based on a discrete injection of cosmic rays by point-like sources at given times. This can be viewed as a refinement of the usual assumption of a continuous and smooth injection of cosmic-rays. This assumption may break down at the highest energies for a number of reasons: (i) At VHE, the cosmic-ray energy density injected by sources per decade in energy is small for injection spectra softer than $2$. (ii) The residence time of cosmic rays in the Galaxy is significantly reduced due to their fast diffusion. (iii) The fraction of sources that are able to accelerate cosmic rays up to the highest energies is still unknown and may be much smaller than at $\rm{TeV}$ energies. As a result, at VHE much less cosmic rays are present in the Galaxy at the same time, allowing for the emergence of local effects resulting from the stochasticity in the locations of the cosmic-ray injection sites.

In order to numerically implement the discreteness of the cosmic-ray injection sites, we adopt the Green's functions formalism and solve analytically the diffusion equation governing the evolution of the cosmic-ray density at each point in the Galaxy. This approach allows to treat the problem from different perspectives in a straightforward way. In particular, we focus our attention on the diffusion regime governing the propagation of cosmic rays and consider two cases: (i) A standard case involving cosmic rays diffusing around their sources with an isotropic and homogeneous diffusion coefficient, similar to the GALPROP assumption. (ii) An alternative case where cosmic rays diffuse slowly in the vicinity of their sources before moving to a fast diffusion regime. This can be achieved by introducing a time dependence for the diffusion coefficient. 

We explore below the impact of the unknown rate of PeVatrons in our Galaxy and of the cosmic-ray propagation mechanism on the number of PeVatrons that are actually observable in our simulations, and on the level of fluctuations of the VHE diffuse gamma-ray background. We find that both the source properties and cosmic-ray propagation mechanism leave imprints on the latter two observables, paving the way for new constraints on the sources.

This paper is organized as follows: In section II, we introduce our model. We derive and present the Green's functions and the other relevant quantities for this study. In section III, we describe the procedure adopted to perform the numerical simulations and our modeling of the Galaxy. In section IV, we display our sky maps, we introduce and define our criterion to identify the detectable sources, we give a theoretical estimate of heir number and discuss the compatibility of our results with observations. In section V, we discuss the impact of the stochasticity on the clumpiness of the gamma-ray flux at VHE and demonstrate its dependence on the number of sources. Finally, we summarize our findings, discuss our model and present our conclusions in section VI. 

\section{Model}
\subsection{Cosmic-ray density and Green's functions}
The propagation of cosmic rays can be described by the full transport equation \cite{generalEquation}. Usually, analytical solutions of this equation are rarely available and one needs to rely on propagation codes, such as GALPROP \cite{galprop} or DRAGON \cite{dragon1, dragon2}, to obtain the cosmic-ray density. This imposes severe limitations on the form of the injection term which needs to be sufficiently simple to keep a reasonable computation time. However, for the newly opened VHE window, most of the contributions to the transport equation become negligible. In this context, the full transport equation can be simplified to become:
\begin{equation}\label{generalDiffusion}
    \frac{\partial n\left(E,\Vec{r},t\right)}{\partial t} - D\left(E, t\right)\Vec{\nabla}^2 n\left(E,\Vec{r},t\right) =N\left(E, \Vec{r}, t\right)
\end{equation}
where $n\left(E,\Vec{r},t\right)$ represents the cosmic-ray density per unit of energy at position $\Vec{r}$ and time $t$, and $D\left(E, t\right)$ is the isotropic and homogeneous diffusion coefficient that only depends on energy and (perhaps) time.

This equation can be solved by means of the Green's functions formalism \cite{blasi, blasi2, pohl2, pohl, mertsch1, mertsch2} leading to a generic solution given by the integral:
\begin{equation}\label{base}
    n(E, \Vec{r}, t) = \int G(E, \Vec{r}, \Vec{r}', t, t')N(E, \Vec{r}', t') d\Vec{r}'dt'
\end{equation}
where the Kernel $G(E, \Vec{r}, \Vec{r}', t, t')$ is the Green's function of the problem and $N(E, \Vec{r}', t')$ represents the source term. For diffusion problems, the Green's function usually takes the form of a Gaussian of the space coordinates, which allows a straightforward definition of the extension $r$ of a source by:
\begin{equation}\label{def}
    r \equiv 1.96\times\sigma
\end{equation}
where $\sigma$ is the standard deviation of the Gaussian and the extension $r$ is, by definition, the radius of a sphere containing $95\%$ of the cosmic rays\footnote{This definition loses all meaning when the cosmic-ray source becomes too old and the escape of cosmic rays becomes important.}.

Assuming (to remain in the conservative side) that the so-called Galactic PeVatrons are supernova remnants (SNRs) injecting PeV cosmic rays as bursts at the very early stages of their lives, the source term can be approximated by:
\begin{equation}\label{spec}
    N\left(E, \Vec{r}, t\right) \equiv N\left(E\right)\delta\left(\Vec{r}-\Vec{r}_s\right)\delta\left(t-t_s\right)
\end{equation}
where $N\left(E\right)$ represents the injection spectrum of the SNR while $\vec{r}_s$ and $t_s$ respectively represent the location of the source and the injection time of its cosmic rays. Inserting (\ref{spec}) in (\ref{base}) allows to express the cosmic-ray density directly by:
\begin{equation}\label{density}
    n(E, \Vec{r}, t) =N\left(E\right) G(E, \Vec{r}, \Vec{r}_s, t, t_s)
\end{equation}
\subsection{Isotropic and homogeneous diffusion}
The standard case of cosmic rays diffusing in the interstellar medium with an isotropic and homogeneous diffusion coefficient can be described assuming a diffusion coefficient reading as \cite{blasi}:
\begin{equation}\label{standarDiffusion}
    D_{Gal}\left(E\right) = 10^{28} D_{28}\left(\frac{R}{3\,\rm{GV}}\right)^{\delta}\, \rm{cm}^2\,\rm{s}^{-1}
\end{equation}
with $D_{28} = 1.33\times\frac{H}{\rm{kpc}}$ for $\delta = 1/3$, corresponding to a Kolmogorov turbulence of the magnetic field. This form, also assumed in GALPROP, ensures that the boron-to-carbon ratio is satisfied for rigidities $R>3\,\rm{GV}$. It should be noted that other choices for the diffusion coefficient, such as the one presented in \cite{weinrich}, are also possible. We nevertheless do not expect such a choice to impact our results, which are insensitive to minor changes in the diffusion mechanism. The Green's function can be determined by means of Feynman's path integrals after the transformation $t\leftrightarrow -it$ and $D\leftrightarrow\frac{\hbar^2}{2m}$ \cite{feynman}. It reads as:
\begin{equation}\label{green1}
   n(E, \Vec{r}, \tau) =N\left(E\right) G\left(E,\Vec{r}, \Vec{r}_s, \tau\right)=\frac{N\left(E\right)}{\left(4\pi D_{Gal}\left(E\right)\tau\right)^{3/2}}\exp{\left(-\frac{\left(\Vec{r}-\Vec{r}_s\right)^2}{4D_{Gal}\left( E\right)\tau}\right)}
\end{equation}
where $\tau = t-t_s$.

Identifying the factor $\sigma \equiv\sigma\left(\tau\right) = \sqrt{2D_{Gal}\left(E\right)\tau}$ in eq. (\ref{def}) as the standard deviation of a Gaussian, leads to a source extension $r$ defined by:
\begin{equation}\label{extension}
    r \equiv 1.96\times\sqrt{2D_{Gal}\left(E\right)\tau}
\end{equation}

While using the diffusion coefficient presented in eq. (\ref{standarDiffusion}) is sufficient for an adequate modeling of the diffuse gamma-ray emission, investigating the detectability of sources is more delicate and requires adjusting equation (\ref{standarDiffusion}). Indeed, at $\rm{PeV}$ energies, eq. (\ref{standarDiffusion}) leads to a diffusion speed bigger than the speed of light for young sources. This behavior, which is expected at all energies, is due to the fact that for length scales that are very short compared to the mean free path of the diffusing particles, the diffusion approximation is not yet valid. The propagation of particles around the source depends on the specific geometry of the magnetic field. At relatively low energies, this phase is very short and can be neglected. However, because of the increase of the mean free path of particles with increasing energy, the importance of this phase grows significantly with energy. At a $\rm{PeV}$, the timescale of this phase coincides with the time window for which cosmic-ray sources are detectable leading to the disappearance of many sources that should still be visible (see section IV).

To cure this problem, we propose a phenomenological diffusion coefficient $D_{p}\left(E,t\right)$ parameterized to satisfy the following two conditions:

\begin{equation}
    D_{p}\left(E,\tau\right) \simeq D_{Gal}\left(E\right)
\end{equation}
for $\tau\rightarrow\infty$ and:
\begin{equation}
    r\left(E,\tau\right) \simeq \frac{c}{2}\tau
\end{equation}
for $\tau\rightarrow0$. 

\begin{figure}[h!]
    \centering
    \includegraphics[scale = 0.507]{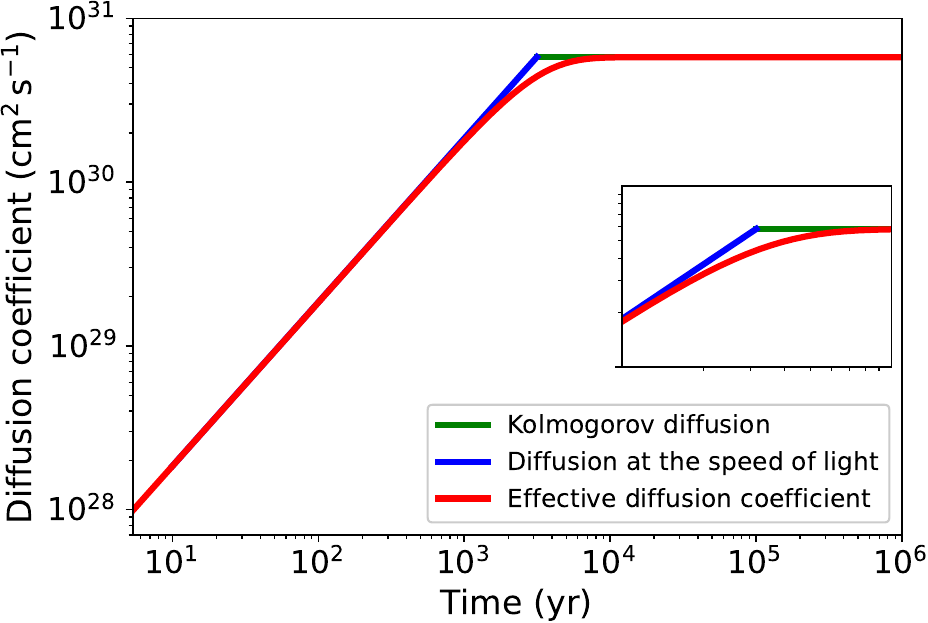}
    \caption{Effective diffusion coefficient (after adjustment) as a function of the source age for $2\,\rm{PV}$ cosmic rays.}
    \label{fig:diff_coeff}
\end{figure}

These two conditions ensure that the size of a source does not grow faster than the speed of light when it is very young and that the diffusion coefficient converges to the standard value after the diffusion speed becomes smaller than the speed of light (see figure \ref{fig:diff_coeff}). It can be shown that these two conditions are satisfied for the transformed diffusion coefficient:
\begin{equation}\label{effective_diffusion}
    D_{Gal}\left(E\right) \rightarrow D_{p}\left(E, \tau\right) = D_{Gal}\left(E\right) \times\tanh{\left(\frac{c^2\tau}{4\times1.96^2D_{Gal}\left(E\right)}\right)}
\end{equation}

In this case, the Green's function and the extension are obtained by applying the transformation:
\begin{equation}
    D_{Gal}\left(E\right)\tau\rightarrow\frac{4\times1.96^2D^2_{Gal}\left(E\right)}{c^2}\times\ln{\left(\cosh{\left(\frac{c^2\tau}{4\times1.96^2D_{Gal}\left(E\right)}\right)}\right)}
\end{equation}
to eq. (\ref{green1}) and eq. (\ref{extension}).

\subsection{Inhomogeneous diffusion}
Recent observations of slow diffusion regions around cosmic-ray sources, such as Geminga and Monogem or the Cygnus cocoon  \cite{geminga, cocoon}, tend to favor the scenario in which cosmic-ray sources would have in their vicinity a suppressed diffusion coefficient, much smaller than that inferred from the boron-to-carbon and other secondary-to-primary ratios. Several alternative models assuming a different value of the diffusion coefficient inside and outside the disc \cite{twozonehalo} or a slow diffusion region around the sources \cite{twozonessphere} have already been proposed. In this regard, we also investigate the imprint of such models on the Galactic VHE gamma-ray flux.

In order to investigate the imprint that slow diffusion regions around the sources would have on the Galactic VHE gamma-ray flux, we assume a time-dependent diffusion coefficient which will act as a proxy for the inhomogeneous diffusion, rather than directly using an inhomogeneous diffusion coefficient, which is more challenging to implement. A time-dependent diffusion coefficient mimicking inhomogeneous diffusion around the sources can be obtained after the transformation:
\begin{equation}\label{timeDependant}
    D_{Gal}\left(E\right) \rightarrow D_{start}\left(\frac{\alpha + 2}{2} + \frac{\alpha}{2}\tanh{\left(\beta\left(\tau-\tau_{change}\right)\right)}\right)
\end{equation}
where $D_{start}$ is the diffusion coefficient seen by the cosmic rays in the slow diffusion regime (just after injection), $\alpha = D_{end}/D_{start}-1$, $D_{end}$ is the final diffusion coefficient perceived by the cosmic rays in the fast diffusion regime and is given by eq. (\ref{standarDiffusion}), $\tau_{change}$ is the age at which the change in the diffusion coefficient is observed, and $\beta$ is the pace of the increase. 

This approach takes advantage of the relationship between the radius of the source and the elapsed time after injection. Shortly after injection, while cosmic rays are still confined close to their source, they experience a slow diffusion dictated by the initial value of the diffusion coefficient which is very small. Subsequently, as they are moving away from their source, while time is passing, they feel a growing diffusion coefficient which will continue to increase, until it reaches its steady value at the time when the cosmic rays leave the slow diffusion sphere.

The cosmic-ray density and the source extension are given by eq. (\ref{green1}) and eq. (\ref{extension}) after the transformation:
\begin{equation}\label{extension2}
    D_{Gal}\left(E\right)\tau\rightarrow D_{start}\left(\frac{\alpha+2}{2}\tau+\frac{\alpha}{2\beta}\left(\ln{\left(\cosh{\left(\beta\tau-\beta\tau_{change}\right)}\right)}+\ln{\left(2\right)}-\beta\tau_{change}\right)\right)
\end{equation}

One can verify that for $\tau \ll\tau_{change}$ the extension takes the form $r \simeq 1.96\times\sqrt{2D_{start}\tau}$ which is identical to the extension defined in eq. (\ref{extension}), while a development for $\tau\gg\tau_{change}$ leads to an extension given by:
\begin{equation}\label{effective}
    r \simeq 1.96 \times\sqrt{2D_{end}\left(\tau-\frac{\alpha}{\alpha+1}\tau_{change}\right)}
\end{equation}
analogous to eq. (\ref{extension}), except that this time, $\tau$ is replaced by an effective age slightly reduced and penalizing the extension for the time spent in the slow diffusion regime. The relative importance of this penalty decreases as the source ages. From this, a definition of the transition radius by means of transition time $\tau_{change}$ arises naturally.

\subsection{The gamma-ray flux}
The computation of the gamma-ray flux observed at Earth is performed through an integration over the line of sight \cite{lipari1}:
\begin{equation}
    \phi_{\gamma}\left(E_{\gamma}, l, b\right) = \int_{0}^{t_{max}}q_{\gamma}\left(E_{\gamma}, l, b, t\right)e^{-\tau\left(E_{\gamma}, l, b, t\right)}dt
\end{equation}
where $q_{\gamma}\left[E_{\gamma}, \vec{x}\right]$ represents the production rate of gamma-rays of energy $E_{\gamma}$ at position $\vec{x}$ and $\tau\left(E_{\gamma}, l, b, t\right)$ represents the optical depth for gamma-rays, accounting for the absorption effects.

At a few hundreds of $\rm{TeV}$, most of the absorption experienced by photons is due either to the cosmic microwave background (CMB) through $\gamma\gamma$ interactions or to the dust present in the interstellar medium. The CMB can be considered to be isotropic and its impact can be modeled by an exponential attenuation factor involving an isotropic optical depth only depending on the gamma-ray energy and the propagation distance \cite{lipari2}. On the contrary, the dust has a non trivial distribution and induces anisotropic absorption effects which may add up to the local effects intrinsically induced by the distribution of cosmic rays. Since the absorption due to the dust is less important than that due to the CMB at a few hundreds of $\rm{TeV}$, we chose to negelect the former.

For the computation of the production rate of gamma-rays, we also neglect a potential leptonic contribution, as it is expected to represent less than $10\%$ of the total diffuse emission at VHE \cite{lipari1}. At a few hundreds of $\rm{TeV}$, the photons from leptonic origin are essentially due to the inverse Compton scattering of CMB photons by $\rm{PeV}$ electrons and positrons whose cooling time is extremely short, preventing them from contributing far from their sources. Therefore, leptons can only contribute to the diffuse background in the form of unresolved leptonic sources which are not relevant for this work. In this context, the VHE gamma-ray emission is hadronic and the determination of the production rate only requires the knowledge of the cosmic-ray density defined previously and a map of the interstellar gas density that we borrow from ref. \cite{lipari1}. This map only reproduces the large scale structure of the interstellar gas and does not account for small scale clumps due to the presence of local structures. This allows to avoid, once again, the appearance of local effects that are not due to the cosmic-ray density. Finally, in order to compute the production rate of cosmic rays, we use the analytical expression \cite{gamma}:
\begin{equation}\label{prod}
    q(E_{\gamma}, \Vec{r}) \equiv \frac{d N_{\gamma}}{dE_{\gamma}} = c n_{H}\left(\Vec{r}\right) \int_{E_{\gamma}}^{E_{max}} \sigma_{inel}\left(E_p\right)n\left(E_p, \Vec{r}, t\right)F_{\gamma}\left(\frac{E_{\gamma}}{E_{p}}, E_p\right)\frac{dE_p}{E_p}
\end{equation}
where the inelastic cross section of the $p-p$ interaction $\sigma_{inel}\left(E_p\right)$ and the energy spectrum of the photons emitted $F_{\gamma}\left(\frac{E_{\gamma}}{E_{p}}, E_p\right)$, are given in ref. \cite{gamma}. We then account for the contribution from heavier nuclei by introducing a nuclear enhancement factor $\epsilon \sim 2.1$ \cite{enhancement}. The analytical expression (\ref{prod}), derived from simulations with the SIBYLL code \cite{sibyll}, is suited for the energy range considered in this work and allows to take into account the contribution to the gamma-ray flux coming from the decay of neutral pions $\pi^0$ as well as $\eta$ mesons while keeping a reasonable computation time.

\section{Numerical setup}
We model the Galaxy as a cylinder of infinite radius and height $H = 5\,\rm{kpc}$ corresponding to the height of the Galactic halo, and we consider that all cosmic-ray sources and most of the interstellar gas are contained in the Galactic disc, whose height is set to $0.2\,\rm{kpc}$. For this geometry with the boundary conditions $n\left(E,\Vec{\rho}, \pm H,t\right) = 0$ and $D\left(E\right)\left(\frac{\partial n\left(E,\Vec{r},t\right)}{\partial z}\right)_{z=\pm H}$ representing the escape flux, we take the solution of \cite{blasi} and then approximate it using the expression given in the appendix of ref. \cite{pohl}. We generate lists of sources of random ages at a rate of $2$ sources per century \cite{rate} at random locations following the surface distribution \cite{param1}:
\begin{equation}
    n\left(r\right) \propto\left(\frac{r}{R_{\odot}}\right)^{0.7}\exp{\left(-3.5\frac{r-R_{\odot}}{R_{\odot}}\right)}
\end{equation}
where $R_{\odot}\simeq 8.5\,\rm{kpc}$ represents the distance of the sun to the Galactic center, while the $z$-distribution is assumed to be uniform between $-0.2\,\rm{kpc}$ and $0.2\,\rm{kpc}$. However, we consider that only a fraction $\chi_{PeV}$ of all supernovae are PeVatrons and will contribute to the VHE gamma ray emission. We ignore the remaining sources and assume that they will contribute only at smaller energies. For each list of sources, out of which a certain percentage are PeVatrons, we fix the value of the injection index in order to fit the cosmic ray flux at Earth at $\rm{PeV}$ energy. We assume that each PeVatron can accelerate cosmic rays to energies ranging from $E_0=1\,\rm{GeV}$ to $E_{max} = 4\,\rm{PeV}$ and has a conservative energy budget $\varepsilon=10^{51}\,\rm{erg}$ of which $\eta =10\%$ contributes to particle acceleration. We also consider that the injection spectrum of SNRs is a power-law of index $s$ of the energy and that the sources inject cosmic rays as bursts. Thus, the injection spectrum of a given source is given by the extension of the formula presented in \cite{blasi} and reads as:
\begin{equation}
    N\left(E\right) = 
    \begin{cases}
        \frac{(s-2)\eta\varepsilon}{E_{0}^{2}}\frac{1}{\left(1-\left(\frac{E_{max}}{E_0}\right)^{-s+2}\right)}\left(\frac{E}{E_0}\right)^{-s}, & s>2 \\
        \frac{\eta\varepsilon}{ln\left(\frac{E_{max}}{E_{0}}\right)}E^{-2}, & s = 2
    \end{cases}
\end{equation}

\begin{figure}[h!]
    \centering
    \includegraphics[width=0.65\linewidth]{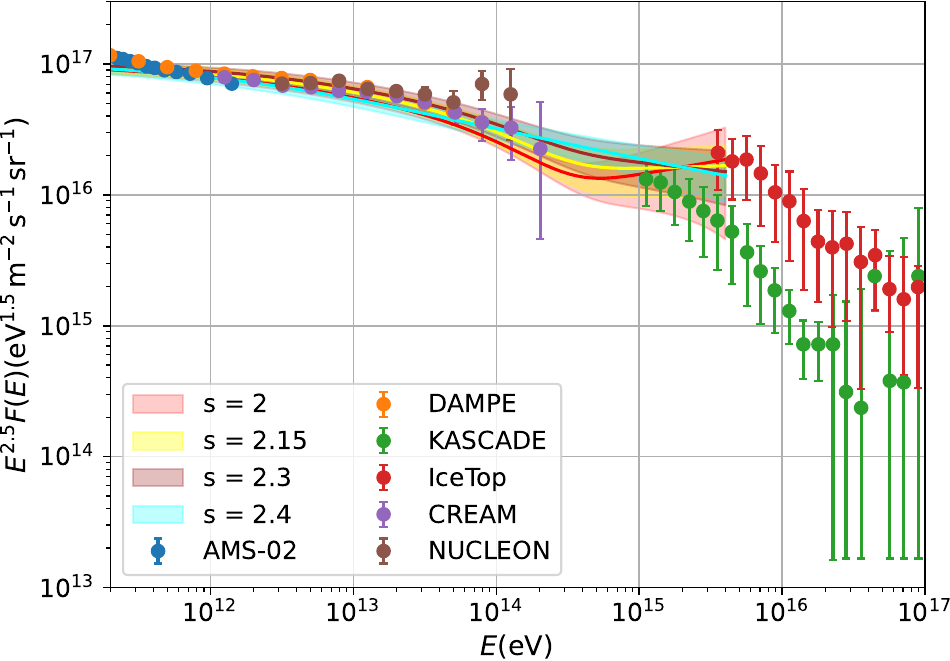}
    \caption{Fit of the cosmic-ray flux assuming different injection indices for the population of PeVatron SNRs. The injection index $s = 2$ corresponds to a fraction $\chi_{PeV} = 2.1\%$ of SNRs that are PeVatrons, $s = 2.15$ corresponds to $\chi_{PeV} =6.7\%$, $s = 2.3$ to $\chi_{PeV} =31.7\%$ and $s = 2.4$ to $\chi_{PeV} =100\%$. For each case where PeVatron SNRs have an injection index harder than $2.4$, the rest of the SNRs population has been assumed to have an injection index of $2.4$, a maximum energy of $200\,\rm{TeV}$ and an exponential cutoff with $E_{cut} = 150\,\rm{TeV}$. The data points are from KASCADE \cite{proton_flux}, AMS-02 \cite{ams}, DAMPE \cite{dampe}, CREAM \cite{cream}, NUCLEON \cite{nucleon} and IceTop \cite{icetop}.}
    \label{fig:fit}
\end{figure}

Using these parameters, we first generate a random list of sources including a given percentage of PeVatrons and fix the spectral index accordingly, following the approach of \cite{anis}. Then, we compute the average flux of cosmic rays over $36$ locations all at a distance of $8.5\,\rm{kpc}$ from the Galactic center and iterate the process until generating a list producing an average flux of protons equal to that reported in \cite{proton_flux}. More precisely, we allow the average value of the cosmic-ray flux to vary by less than $0.2\%$ from one simulation to another to ensure the validity of the results presented in section V. It should be noted that, as long as only a subset of all SNRs are given a spectrum harder than $2.4$, the cosmic-ray flux is well-fitted for GeV energies, while in the region between the bump at a few tens of TeV and the knee, the index of the cosmic-ray spectrum is poorly known due to the large error bars, potentially leading to a spectrum harder than below the bump. This is illustrated in figure \ref{fig:fit} which shows the cosmic-ray spectrum derived considering different injection indices\footnote{However, we do not expect any of the considered injection indices to fit the recent gamma-ray data of LHAASO exhibiting a spectral index of $2.99$. This very soft spectrum is likely to be due to a contamination from unresolved leptonic sources whose contribution remains uncertain \cite{me, neutrinos, hooper}.}. Moreover, assuming two populations of sources injecting particles with two different indices, may provide a plausible explanation for the bump at a few tens of TeV, which happens to be close to the maximum energy achievable in type Ia supernovae according to \cite{bell}. However, such an investigation would require a proper fitting of the different parameters, which is beyond the scope of this paper.

After having generated the sources, we propagate the cosmic rays in the galaxy for each of the two diffusion mechanisms described previously and compute the resulting gamma-ray flux. The standard case of a homogeneous diffusion coefficient is described by eq. (\ref{effective_diffusion}) assuming a Kolmogorov turbulence of the magnetic field, while the case of the inhomogeneous diffusion is described by eq. (\ref{timeDependant}), where we set the value of the diffusion coefficient after cosmic-ray injection to $D_{start} = 10^{28}\,\rm{cm}^2\,\rm{s}^{-1}$, the age at which the value of the diffusion coefficient changes to $\tau_{change} \simeq 10\,\rm{kyr}$ (corresponding to a radius $r_{change}\simeq50\,\rm{pc}$) and the pace of increase to $\beta=100\,\rm{yr}^{-1}$. We have chosen these parameters in order to ensure that the time that cosmic rays spend close to their sources in the slow diffusion regime remains negligible compared to their residence time in the Galaxy. This, in turn, guarantees that cosmic rays do not accumulate too much boron and therefore preserves the boron-to-carbon ratio\footnote{It should be noted that, so far no measurement of the boron-to-carbon  ratio at $\rm{PeV}$ energies has been made and that the constraint that we aim to satisfy here is an extrapolation of the $\rm{GeV}-\rm{TeV}$ data.}. Another interesting situation not considered in this work would be to assume that cosmic rays accumulate a non negligible part of their boron close to their sources \cite{grammage} by confining them longer in the vicinity of their sources before allowing them to transition to a very fast diffusion regime.

\section{Gamma-ray flux and cosmic-ray source detectability}
Using our model, we have generated, for the two diffusion processes, VHE gamma-ray maps of our Galaxy that include the contribution of the diffuse background, as well as that of the point-like and extended sources. We present our results related to the homogeneous diffusion mechanism (first case) in figure \ref{fig:skymap1} and those related to the inhomogeneous diffusion mechanism (second case) in figure \ref{fig:skymap2}. In both figures, we have displayed the gamma-ray flux measured at Earth in the direction $\left(l, b\right)$ for different fractions of the total number of sources that are PeVatrons, with the corresponding spectral index.
\subsection{Homogeneous diffusion}

\begin{figure}[h!]
    \centering
    \includegraphics[scale = 0.49]{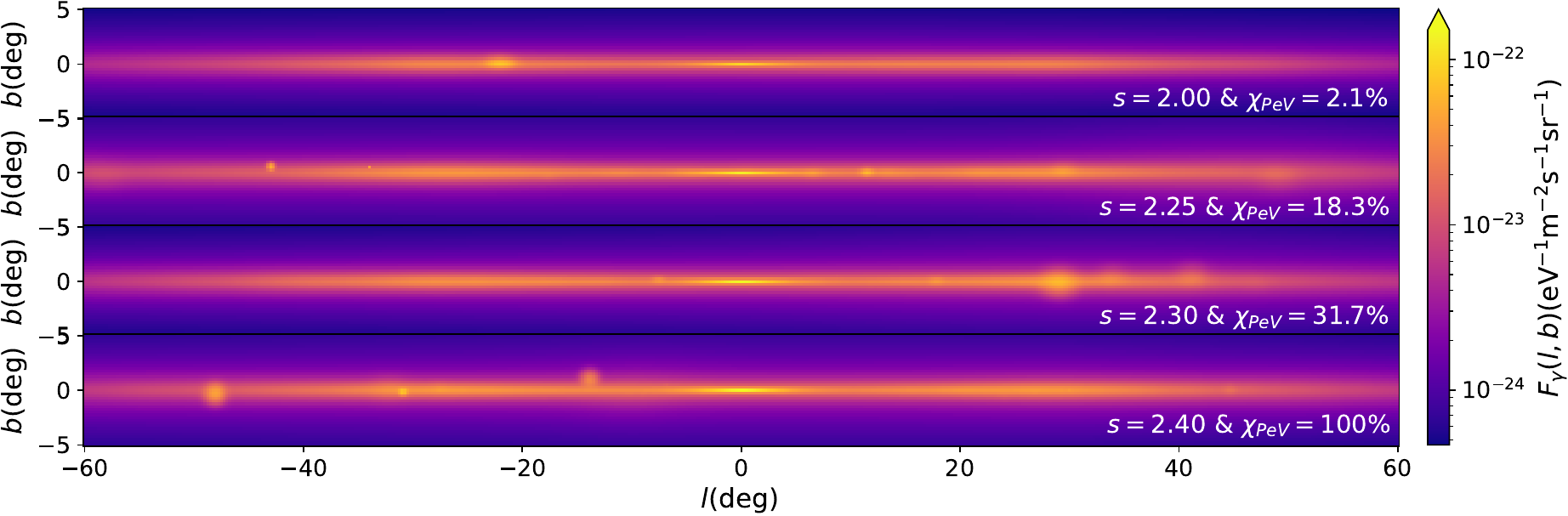}
    \caption{Gamma-ray flux at $E = 200\,\rm{TeV}$ as a function of the Galactic longitude $-60^{\circ}\leq l\leq60^{\circ}$ and the Galactic latitude $-5^{\circ}\leq b\leq5^{\circ}$ in the case of a standard isotropic and homogeneous diffusion coefficient represented by eq. (\ref{effective_diffusion}) for different percentages of sources that are PeVatrons, with the corresponding spectral index.}
    \label{fig:skymap1}
\end{figure}

In figure \ref{fig:skymap1}, we generally observe few visible sources for all percentages of SNRs that are PeVatrons. Of course, there is still a correlation between the number of PeVatrons in the Galaxy and the number of them that are actually visible, in the sense that it is difficult to observe visible sources in the case where only $2.0\%$ of them are PeVatrons, or that the higher the number of PeVatrons is the more chances we have to observe some of them. However, for mainly two reasons, this correlation is weakened at VHE if a homogeneous diffusion in the interstellar medium is assumed. 

First, at $\rm{PeV}$ energy, cosmic rays diffuse so fast (their diffusion coefficient is $10^2$ times bigger than at $\rm{GeV}$ energy) that it becomes extremely difficult for any source to remain physically compact for more than a few hundreds of years after the last $\rm{PeV}$ cosmic rays have been injected. This has the consequence to considerably shorten the time window during which a SNR remains observable as a PeVatron, which if related to the Galactic rate of supernovae $\sim0.01-0.03\,\rm{yr}^{-1}$ requires us to be in a very favorable situation to be able to observe it. It should also be noted that in the absence of the correction introduced for the diffusion coefficient in section II, the lack of visible PeVatron SNRs is even more pronounced, as the very small time window during which the sources are visible happens to coincide with the time window when the diffusion approximation is not valid, leading the sources to disappear even faster. 

\begin{figure}[h!]
    \centering
    \includegraphics[scale=0.49]{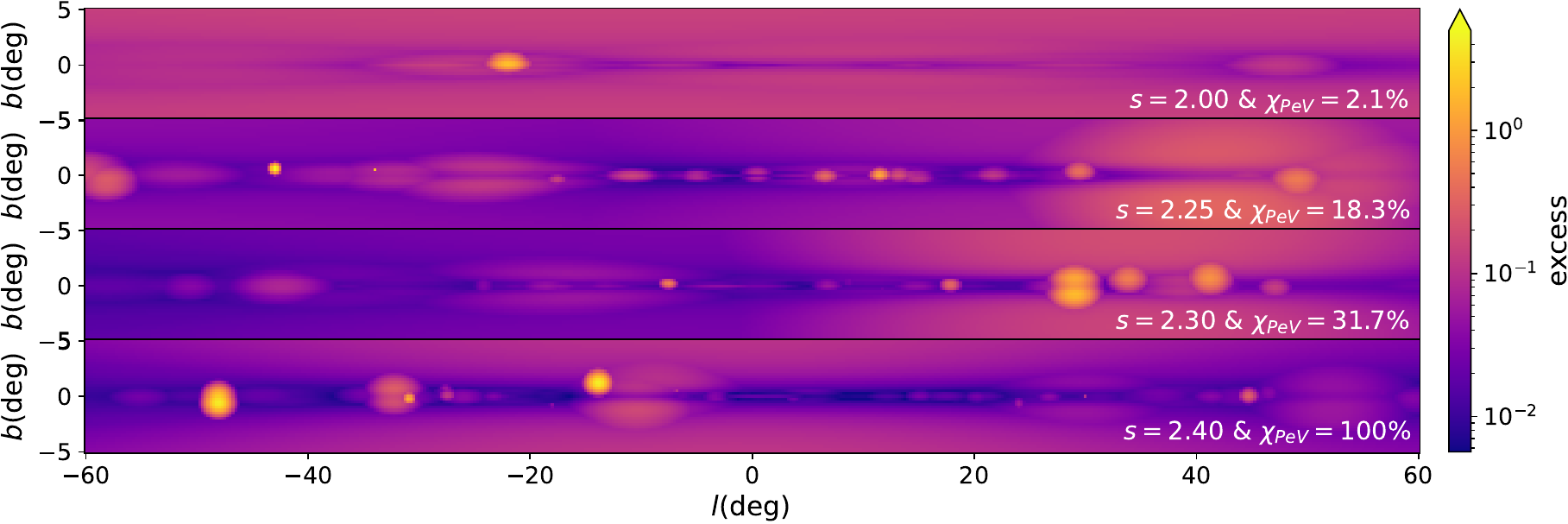}
    \caption{Relative contribution of the brightest source compared to its background associated to the sky maps of figure \ref{fig:skymap1}.}
    \label{fig:significance1}
\end{figure}

The second reason for this lack of visible PeVatrons is the fact that gamma-ray observations are a projection of our 3D Galaxy into a 2D sky map. In this case, even a SNR that is sufficiently young and compact to be observed still needs to compete with all the emissions coming from the diffuse background and other sources along its line of sight. This can be visualized in figure \ref{fig:significance1} displaying the excess of the brightest source of each pixel compared to its background made of the diffuse emission and the emission of other sources. This excess is defined by:
\begin{equation}\label{criterion}
    excess \equiv \frac{brightest}{background=(total - brightest)}
\end{equation}
In figure \ref{fig:significance1}, we see that several more sources are present compared to the intensity map in figure \ref{fig:skymap1} and have a proper gamma-ray emission, but, this emission is overtaken by their background. We also point the fact that here, the correlation between the number of PeVatron SNRs and the number of observable PeVatron SNRs is more obvious. This shows that in general, in order for a source to be visible, it first needs to be sufficiently young and compact, and it also should be in a location where the gamma-ray emissivity is high, but at the same time, it needs a low gamma-ray emissivity elsewhere along the lines of sight passing by its location. In other words, a source whose position coincides with the Galactic ridge but is far from the Galactic center (where the gas density is higher) has very few chances to be visible. In summary, in the scenario presented here, the ideal target for the detection of a PeVatron SNR, would be one very young and compact source, not too far from Earth and, ideally, a little bit offset from the Galactic plane so that it becomes easier for it to overcome its background.

These conclusions are supported by the latest LHAASO data. The results presented here go along with the evident lack of SNR PeVatrons in the LHAASO catalog \cite{catalog}. In addition, this scenario is also supported by the fact that the contribution of sources to the Galactic neutrino flux above $100\,\rm{TeV}$ seems to be small \cite{neutrinos}.

\subsection{Inhomogeneous diffusion}
\begin{figure}[h!]
    \centering
    \includegraphics[scale=0.49]{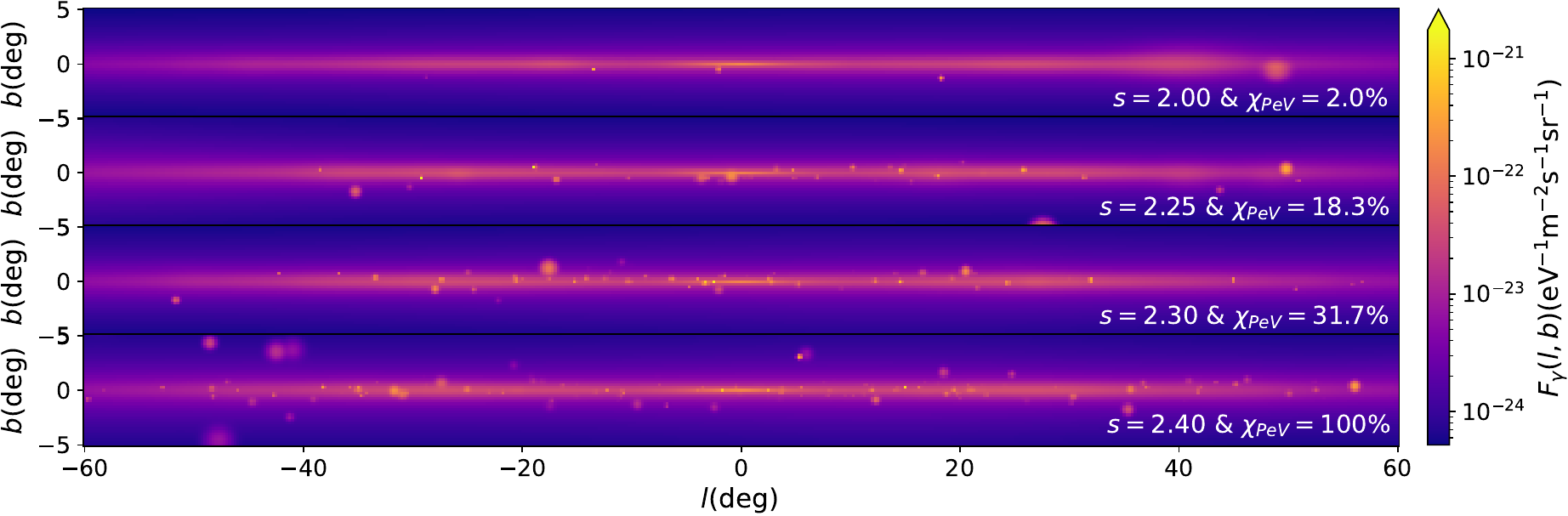}
    \caption{Gamma-ray flux at $E = 200\,\rm{TeV}$ as a function of the Galactic longitude $-60^{\circ}\leq l\leq60^{\circ}$ and the Galactic latitude $5^{\circ}\leq b\leq5^{\circ}$ in the case of an inhomogeneous diffusion coefficient for different percentages of sources that are PeVatrons, with the corresponding spectral index.}
    \label{fig:skymap2}
\end{figure}

In figure \ref{fig:skymap2}, unlike figure \ref{fig:skymap1}, we observe a very strong correlation between the number of visible sources and the number of PeVatron SNRs in the simulation. Here, we see that forcing cosmic rays to diffuse slowly while they are confined inside a sphere of radius $50\,\rm{pc}$ around their sources leads them to remain sufficiently compact for more than $10\,\rm{kyr}$, which related to the Galactic supernova rate $\sim0.01-0.03\,\rm{yr}^{-1}$ gives, this time, a sample of sources sufficiently large for the statistical quantities and relationships to recover their meaning.

\begin{figure}[h!]
    \centering
    \includegraphics[scale=0.49]{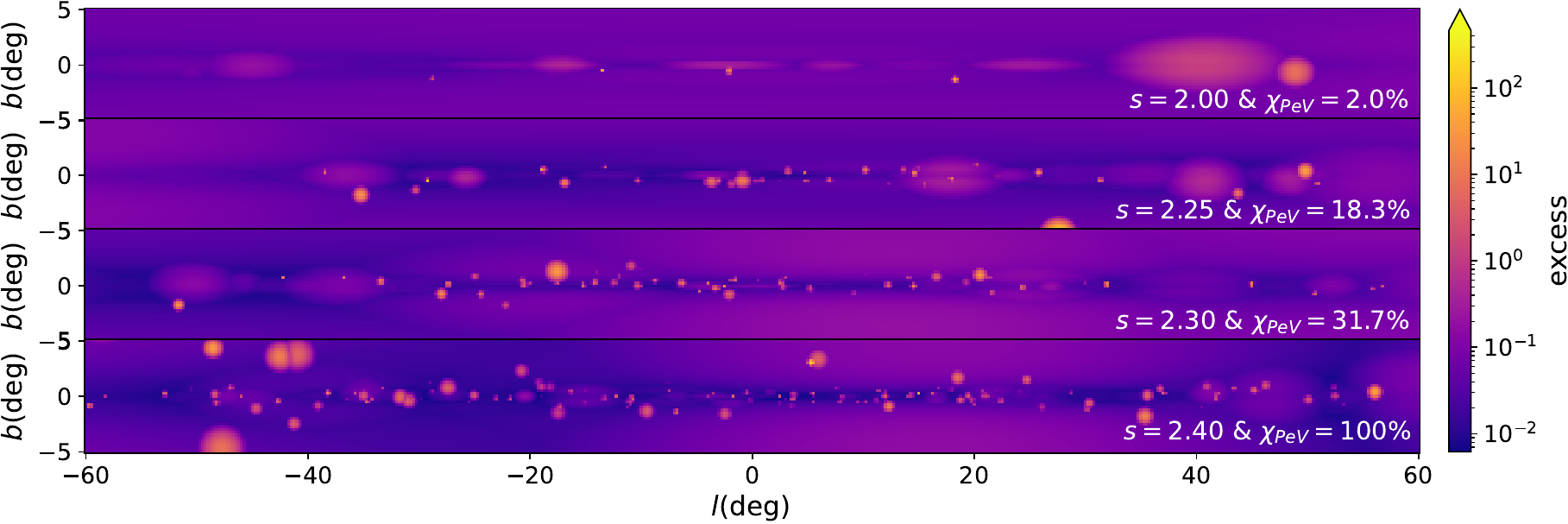}
    \caption{Relative contribution of the brightest source compared to its background associated to the sky maps of figure \ref{fig:skymap2}.}
    \label{fig:significance2}
\end{figure}

Moreover, figure \ref{fig:significance2} representing the excess defined by eq. (\ref{criterion}) confirms that allowing the existence of a sphere of slow diffusion around the sources only impacts the young sources (that are in the slow diffusion regime) and leaves a negligible imprint on the population of old or relatively old sources, which end up behaving like in the case of a homogeneous diffusion after transitioning to the fast diffusion regime. This is evident from the relatively extended regions of small but visible source excess over the background similar to those in figure \ref{fig:significance1}. 

Interestingly, this shows that in this scenario too, even young and compact SNRs do not always manage to overcome their background, even though they benefit from a higher cosmic-ray density in their surroundings for longer durations. This is evident from the compact spots present in figure \ref{fig:significance2} but not in figure \ref{fig:skymap2}. We conclude that in the case of inhomogeneous diffusion the main restriction preventing the sources from being detectable is their inability to overcome their background when it is too bright, while the restriction on their age is significantly softened.

\subsection{Constraints on the number of PeVatron SNRs in the Galaxy}
\begin{figure}[h!]
    \centering
    \includegraphics[scale = 0.569]{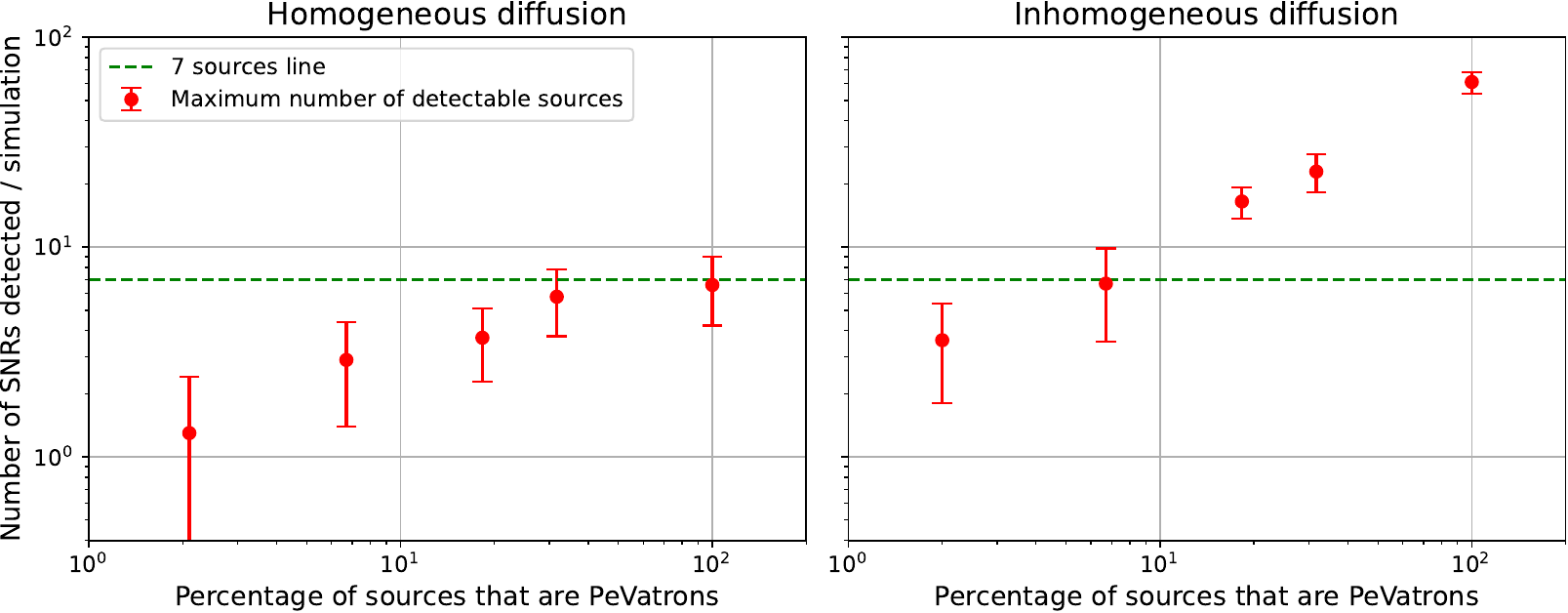}
    \caption{Average maximum number of PeVatron SNRs detected per simulation as a function of their percentage in the Galaxy, with the $1\sigma$ statistical uncertainty.}
    \label{fig:detected_sources}
\end{figure}

In order to constrain the number of PeVatron SNRs in our Galaxy, we have run a total of $10$ simulations per percentage of PeVatron SNRs in the Galaxy and for each diffusion mechanism. After obtaining the intensity map in $\left(l, b\right)$ for each realization, we apply the following procedure: First, we select the brightest source of each pixel for $10^{\circ}\leq l \leq240^{\circ}$ and $\left|b\right|\leq5^{\circ}$. This roughly corresponds to the region where LHAASO reported most of its Galactic sources in its catalog \cite{catalog}. We then compare the excess of the selected sources to their background and require that it is bigger than $13.25\%$ if $10^{\circ}\leq l \l<125^{\circ}$ and bigger than $39.96\%$ if $125^{\circ}< l \leq240^{\circ}$, which correspond to the uncertainties in the value of the diffuse background measured by LHAASO for respectively the inner and outer Galaxy in \cite{lhaaso1}. In addition, we apply cuts on the size of the sources, requiring that their angular extension\footnote{Here, we define the angular extension as the angular radius of the region over which the source is detected rather than taking the definition of section II.} remains smaller than $5^{\circ}$ and that their asymmetry (the ratio of their big semi-axis to their small semi-axis), if they are extended (angular extension bigger than $2.5^{\circ}$), remains smaller than $4$. We have set this additional restriction to remain consistent with observations. In addition, it should be noted that the fact that our simulations produce asymmetric sources is not due to the diffusion mechanism. This is due to the fact that there is no sudden drop in the target gas density with longitude, while this same density drops quickly with latitude. Finally, we assume that all sources benefit from a form of enhancement of their emission owing to the presence of a nearby over-density of interstellar gas in the form of a clump inside the shell of the SNR or the presence of a nearby molecular cloud. This choice is motivated by the fact that the vast majority of supernovae occur close to a molecular cloud and that most detected SNRs have their gamma-ray spectra fitted with a gas density much higher than the canonical $1\,\rm{cm}^{-3}$. In that sense, our result should not be viewed as a prediction of the number of sources that LHAASO will detect, but as constraints from observations of what is realistic in terms of number of cosmic-ray sources and diffusion mechanisms in the Galaxy. We present our results in figure \ref{fig:detected_sources} which displays the maximum number of detected SNRs (red points) and the seven sources line (dashed line in green) that we will use a reference value\footnote{This choice is arbitrary and roughly corresponds to the number of new sources detected by LHAASO and possibly associated with an SNR. The incoming discussion does not depend on the value chosen here.} for the incoming discussion.

We see that the left (right) panel of figure \ref{fig:detected_sources} reflects the trend already discussed in this section for the homogeneous (inhomogeneous) diffusion. The number of detectable SNRs barely increases with the increase of their number in the Galaxy in the homogeneous case, while the correlation is very clear in the case of inhomogeneous diffusion. 

More quantitatively, in the case of homogeneous diffusion, it is very unlikely that there are more than $10$ SNRs in the field of view of LHAASO that would satisfy the physical requirements to have a chance of being observed. Moreover, the number of potentially detectable sources is usually very small and relatively similar for all percentages of PeVatron SNRs in the Galaxy. This makes it very challenging to infer the actual rate of PeVatrons.

In the inhomogeneous diffusion scenario, on the other hand, if a number of seven detected sources is assumed, constraints can already be set on the number of PeVatron SNRs in the Galaxy. In particular, the cases with more than $\sim30\%$ of PeVatron SNRs are shown to lead to the appearance of many more sources than the actual number of possible SNR candidates. In fact, any number of detected sources gives valuable information about the actual number of sources in the Galaxy. If more than seven sources are detected in the future, this will rule out the cases with small numbers of PeVatron SNRs in the Galaxy, while a very small number of detected sources will be a strong evidence of their scarcity.

However, we also see that the approach which consists in discriminating based on the number of detected sources still leaves some degeneracy. The case of inhomogeneous diffusion with small percentages of sources $\lesssim18\%$ leads to the same number of observable sources as in a big part of the entire range in the homogeneous diffusion case. This makes the disentanglement between the two situations, solely based on the number of sources that are detected, impossible.

Finally, we note that in all cases the number of detectable sources tends to increase when smaller halo sizes are considered. Nevertheless, our previous conclusions remain unchanged for different halos sizes. More details on the impact of the halo can be found in the appendix.

\section{The diffuse gamma-ray emission}
In this section, we aim to describe the morphology of the VHE gamma-ray emission and that of its diffuse component. In this regard, we compute and analyze the gamma-ray flux as a function of the Galactic longitude for the same lists of sources as in the previous section.
\subsection{The total gamma-ray emission}

\begin{figure}[h!]
    \centering
    \includegraphics[scale = 0.79]{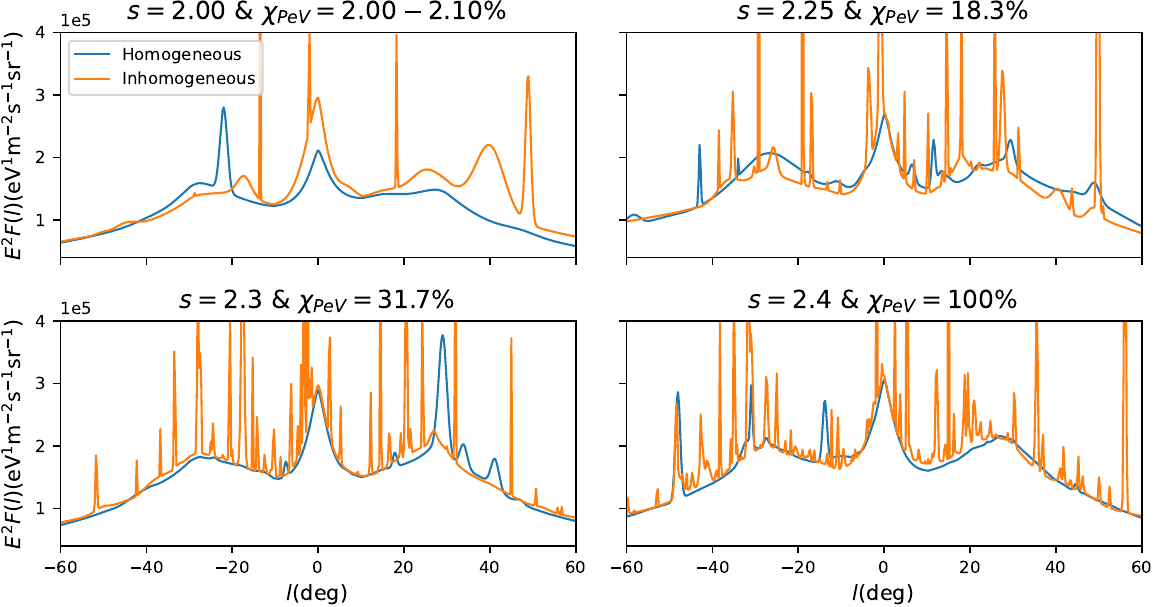}
    \caption{Gamma-ray flux as a function of $l$ for $|b|\leq5^{\circ}$ for different percentages of sources that are PeVatrons ($\chi_{PeV}$) in the homogeneous, and inhomogeneous diffusion mechanisms at $E_{\gamma} = 200\,\rm{TeV}$.}
    \label{fig:brutFlux}
\end{figure}

In figure \ref{fig:brutFlux}, we plot the gamma-ray flux integrated over $|b|\leq5^{\circ}$ as a function of the Galactic longitude for both diffusion mechanisms. It represents the sum of a diffuse component, that is somewhat similar for the two diffusion mechanisms and for the different percentages of sources that are PeVatrons, and a source component appearing as spikes on top of the diffuse background. As expected, We see many more spikes going higher in the case of inhomogeneous diffusion around the sources. This is due to the presence of many more sources whose (very young) effective ages allow to remain very compact and have a very high gamma-ray flux. 

Moreover, in the inhomogeneous diffusion case, higher numbers of sources generally lead to lower spikes on top of the background which remains constant. This is a consequence of the fact that we use the injection index as a free parameter to balance the change in the percentage of PeVatron SNRs and fit the cosmic-ray flux at Earth. In this situation, the simulations with less sources having harder injection spectra will exhibit higher spikes, since these sources inject a higher energy density in the VHE range, while the simulations with many sources having a softer spectrum will exhibit lower spikes, since there is a smaller energy density in the last decades in energy. This should also apply to the homogeneous diffusion case although it is not visible due to the small number of detectable sources.

In addition to the spikes, we also observe shape variations that depend on the fraction of PeVatron SNRs for both diffusion mechanisms. These variations are the most extreme when the number of sources is the lowest (upper left panel of figure \ref{fig:brutFlux}), in which case we may even see an asymmetry between the left and the right sides of the plot. We attribute these fluctuations, that tend to decrease as the number of sources increases, to the stochasticity in the source locations. It should be noted however that these fluctuations never entirely disappear. It has been shown in \cite{stall}, where the authors also conducted Monte-Carlo simulations, that even with $100\%$ of SNRs being PeVatrons, the statistical fluctuations from one simulation to another remain higher than at $\rm{GeV}$ energies.
\subsection{The diffuse gamma-ray background}
\begin{figure}
    \centering
    \includegraphics[scale=0.79]{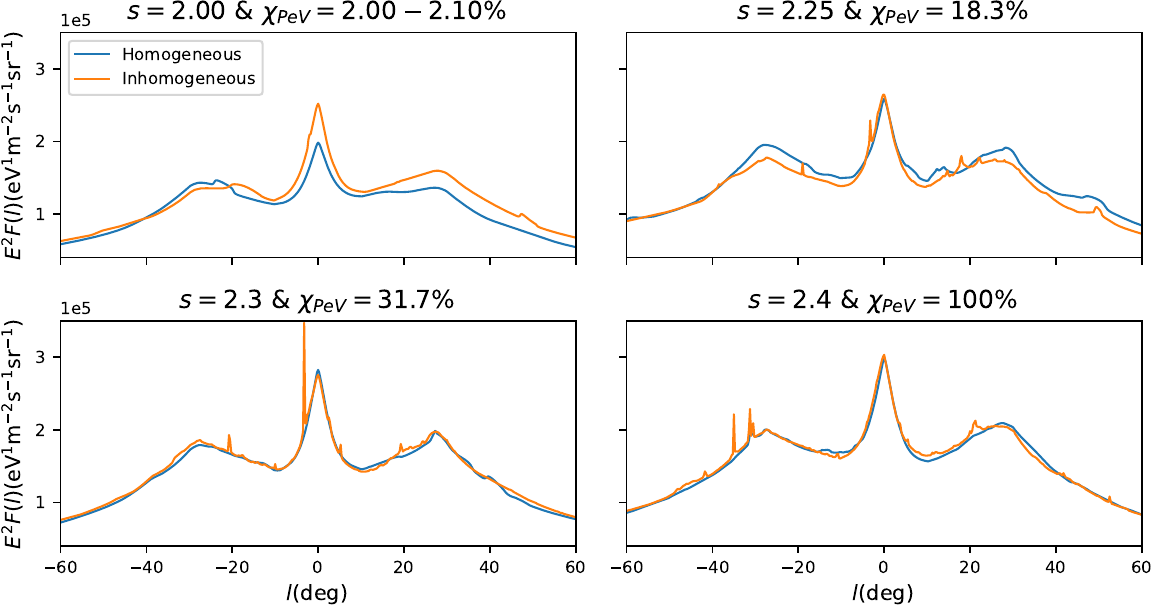}
    \caption{Gamma-ray flux as a function of $l$ for $|b|\leq5^{\circ}$ with the brightest source of each pixel masked for different percentages of sources that are PeVatrons ($\chi_{PeV}$) in the homogeneous, and inhomogeneous diffusion mechanisms at $E_{\gamma} = 200\,\rm{TeV}$.}
    \label{fig:trueFlux}
\end{figure}

In figure \ref{fig:trueFlux}, we attempt to represent the diffuse component of the total gamma-ray emission. In order to do so, we remove the contribution of the brightest source of each pixel and define the remaining flux as the diffuse background. We acknowledge that this procedure is not perfect, in that it might for instance fail to remove all the source contribution from pixels hosting two (or more) coincident sources, as it is the case around $l\simeq-5^{\circ}$ in the lower left panel of figure \ref{fig:trueFlux}. However, in both of the homogeneous and inhomogeneous cases, this instance is not very frequent and care has been taken in handling it.

We see that the removal procedure has not completely smoothed the appearance of the gamma-ray emission. Like before, shape variations and asymmetries between the left and right sides of the plots (decreasing with the increase in the number of sources) are still visible. This shows that the structure of the background is intrinsically complex. Even after having systematically subtracted the contribution of the brightest source from each pixel, there are still non negligible local contributions. These contributions have two possible origins: Either they come from extended sources that are relatively old and whose flux blends into the true background, or they reflect the existence of some regions of high (low) cosmic-ray density that have hosted more (less) SNRs throughout their history. Although both scenarios are possible and are not exclusive, we argue that the former is dominant as it is statistically more likely to happen. 

In addition to the local deviations appearing in figure \ref{fig:trueFlux}, we can also observe a systematic or global shift of the entire curve from one simulation to another. This is particularly visible in the upper panels of the same figure. We attribute this shift to the fact that at VHE the gamma-ray emission is sensitive to the distribution of cosmic rays (and their sources) in the Galaxy. Even when the average cosmic-ray density does not vary by more than $0.2\%$ from one simulation to another, we can still observe much stronger variations in the gamma-ray flux. As a result, the fluctuations in the gamma-ray flux emission are the outcome (not necessarily the sum) of the two effects combined.

\begin{figure}[h!]
    \centering
    \includegraphics[scale = 0.74]{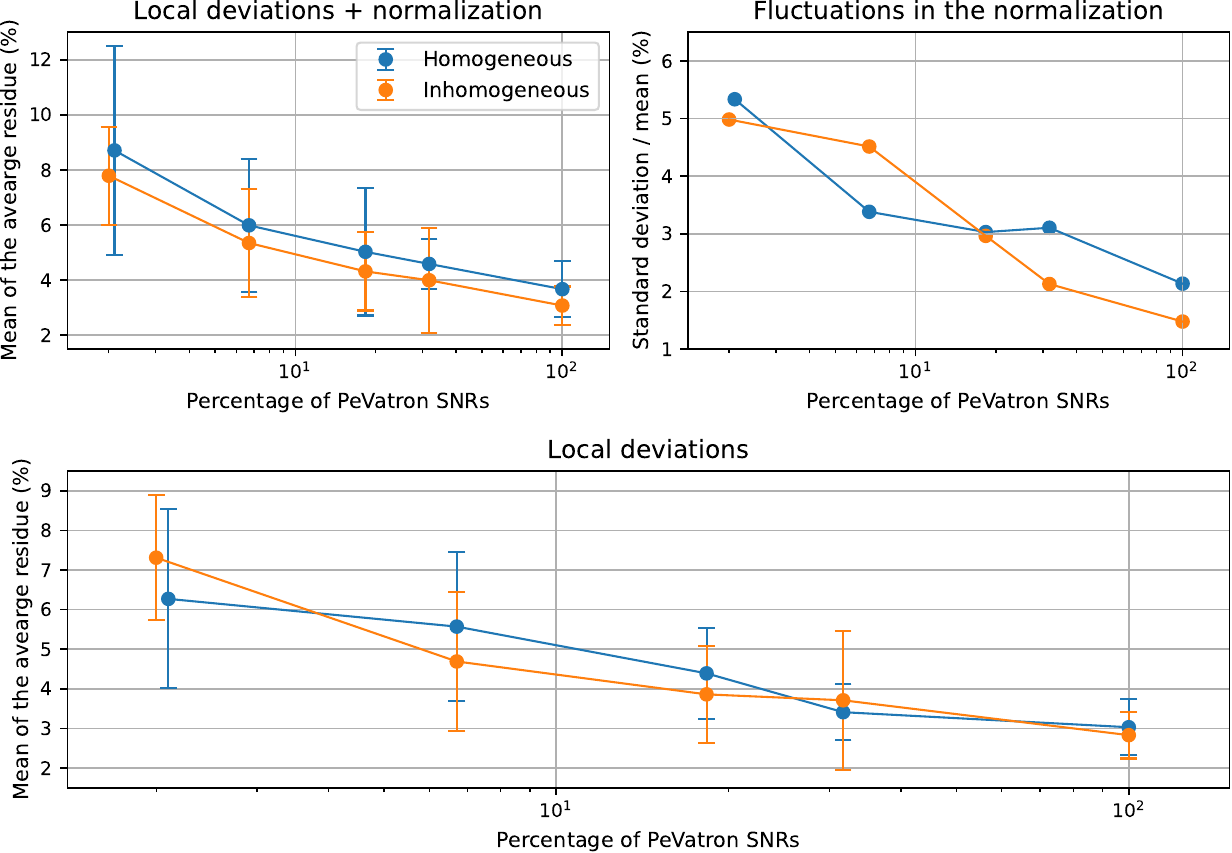}
    \caption{Fluctuations experienced by the VHE diffuse gamma-ray flux at $200\,\rm{TeV}$.}
    \label{fig:fluctuations}
\end{figure}

In order to quantify the impact of the two effects and their dependence on the number of sources in the Galaxy, we analyze the same sets of simulations as in section IV and display the results in figure \ref{fig:fluctuations}. 

The upper left panel quantifies the fluctuations on the gamma-ray flux occurring as a result of the shift in the absolute normalization and the local shape variations. The plot displays the mean, over the ten simulations, of the residue of the gamma-ray flux of figure \ref{fig:trueFlux}, averaged over $-180^{\circ}<l\leq180^{\circ}$, with respect to the median of the realizations. The median is chosen instead of the mean in order to mitigate the effect induced by a source that the procedure introduced previously would fail to remove. For both diffusion mechanisms, we see a strong correlation between the number of sources and the level of fluctuations, as the case with $\sim100\%$ of PeVatron SNRs is disconnected from the case with $\sim2\%$ for which the average residue reaches $\sim17\%$ in the most extreme case\footnote{We only consider the average residue over the entire Galaxy rather than the maximum of local deviations to avoid being mislead by any residual source.}, with a maximum of $\sim37.6\%$ around $l\simeq-150^{\circ}$. For this particular simulation, we have checked by hand that the fluctuation occurs on a large scale and is not due to a point source. However, the fluctuations presented here include the combined effects of the local deviations and the systematic shift in the level of the flux, compared to a reference flux that is impossible to determine experimentally in the absence of a good knowledge of the cosmic-ray density everywhere in the Galaxy. Furthermore, the variations in the normalization of the diffuse flux are not negligible compared to the total fluctuations. This is visible in the upper right panel of figure \ref{fig:fluctuations}, which shows the ratio of the standard deviation and the mean over the ten simulations.

To quantify the intensity of the fluctuations in a measurable way, we show in the lower panel of figure \ref{fig:fluctuations} the same quantity as the upper left panel, except that this time, before computing the residues we normalize each curve so that its average energy coincides with the average energy of the median curve. This way, the fluctuations in the absolute level of the flux are canceled and the remaining residue represents the shape variations. Defined this way, this residue is a measurable quantity allowing to discriminate between cases involving different percentages of sources. Although, in general, the fluctuations solely due to the shape variations are weaker than the fluctuations due to the combined effects, they do not differ in the fact that they occur on relatively large scales (few degrees or tens of degrees), which should make them detectable by experiments. In particular, we find that the simulation that previously led to a $\sim17\%$ average residue leads here to a $\sim12\%$ average residue with a maximum of $\sim28\%$ at the same location as previously. We argue, therefore, that studying the diffuse background can offer a complementary criterion, with the number of detectable sources, to constrain the number of PeVatron SNRs in our Galaxy and the diffusion mechanism of cosmic rays. 

We find that the importance of the fluctuations of the diffuse background tend to increase for smaller halo sizes due to the smaller residence time of cosmic rays in the Galaxy, that scales as $\tau\propto H$. A detailed discussion of this effect is given in the appendix.

\section{Discussion and conclusion}
In this paper, we have presented our study of the VHE sky using a model based on the injection of cosmic rays by a stochastic population of SNRs, and have investigated the behavior of these sources and the morphology of the diffuse gamma-ray background assuming two different scenarios for the diffusion of cosmic rays in the interstellar medium.

We have notably shown that in the case of a double zone model where the diffusion is slow close to the sources it is unlikely that an important fraction of SNRs are PeVatrons, whereas in the standard paradigm involving a single diffusion coefficient equal to the Galactic average the chances to observe PeVatron SNRs is severely reduced due to their inability to remain compact for long times, which goes along with the small number of PeVatron SNRs recently reported by LHAASO \cite{catalog}.

We have also analyzed the morphology of the VHE gamma-ray emission and that of its diffuse component, and found that it is dependent on the birthrate of PeVatron SNRs. In particular, the diffuse background becomes more or less clumpy depending on the number of PeVatrons in the simulation. This behavior is the most extreme for the lowest birthrates of PeVatron SNRs leading to the most important variations in the cosmic-ray sea, while the highest birthrates of PeVatron SNRs lead to a situation closer (but still different) to the $\rm{GeV}$ case. Moreover, assuming a small halo size can further enhance the clumpiness of the diffuse background. In this context, observations of very important fluctuations on the gamma-ray diffuse background may be the signature of a small-sized halo \textbf{(see appendix)}.

We have focused here on the most relevant energy range for the LHAASO and AS$\gamma$ experiments. Our model could nonetheless be extended to other energies. While its extension to higher energies should not reveal any qualitative changes, its extension to lower energies would require some adjustments. At lower energies, we expect sources to be able to accelerate more particles for longer periods of time, which may contribute to smooth the diffuse background and maintain the visibility of the sources for longer periods of time. In addition, it is suspected that at lower energies, the diffuse background suffers from the contribution of gamma rays from unresolved leptonic sources \cite{neutrinos}. This can significantly change the morphology of the measured diffuse background, which would need to be accounted for.

Finally, in this paper, we have not investigated the scenario in which the diffusion would occur in Kraichnan turbulence, because the extremely large mean free path of PeV particles in this turbulence would make the diffusion approximation break down on relevant scales. Moreover, this type of turbulence does not seem to be favored by the boron-to-carbon ratio measurements at lower energies. We have also excluded the situation in which PeVatron SNRs are gathered and occur in star clusters, and the scenario in which the Galactic PeVatrons are star clusters accelerating and injecting cosmic rays continuously for a few millions of years. These two situations would give rise to competing effects in shaping the morphology of the diffuse background, and the final outcome of these effects is uncertain. The space correlation between the sources would reduce the number of injection sites, which would in principle increase the space fluctuations in the cosmic-ray density, while the injection at successive times would in contrast contribute to maintain a more uniform level for the sea of cosmic rays in the Galaxy. Disentangling between these effects would require a dedicated investigation with an elaborate modeling of the star cluster population, which is beyond the scope of the present work.

In conclusion, we have demonstrated that the combined information of the number of observable PeVatron SNRs and the observed level of fluctuations of the VHE gamma-ray background provides a new, valuable help to constrain both the fraction of PeVatron SNRs in the Galaxy and the propagation mechanism of cosmic rays in the interstellar medium. Comparing such predictions with LHAASO and AS$\gamma$ measurements should, at last, help close in on the origin of the bulk of hadronic PeV cosmic rays in our Galaxy.

\acknowledgments
Samy Kaci acknowledges funding from the Chinese Scholarship Council (CSC) and thanks Ramiro Torres-Escobedo for his valuable help and support. This work is supported by the National Natural Science Foundation of China under Grants Nos. 12350610239, and 12393853. This work was supported by the National Center for High-Level Talent Training in Mathematics, Physics, Chemistry, and Biology.


\appendix
\section{Study of the impact of the size of the halo}
\begin{figure}[h!]
    \centering
    \includegraphics[width=1.\linewidth]{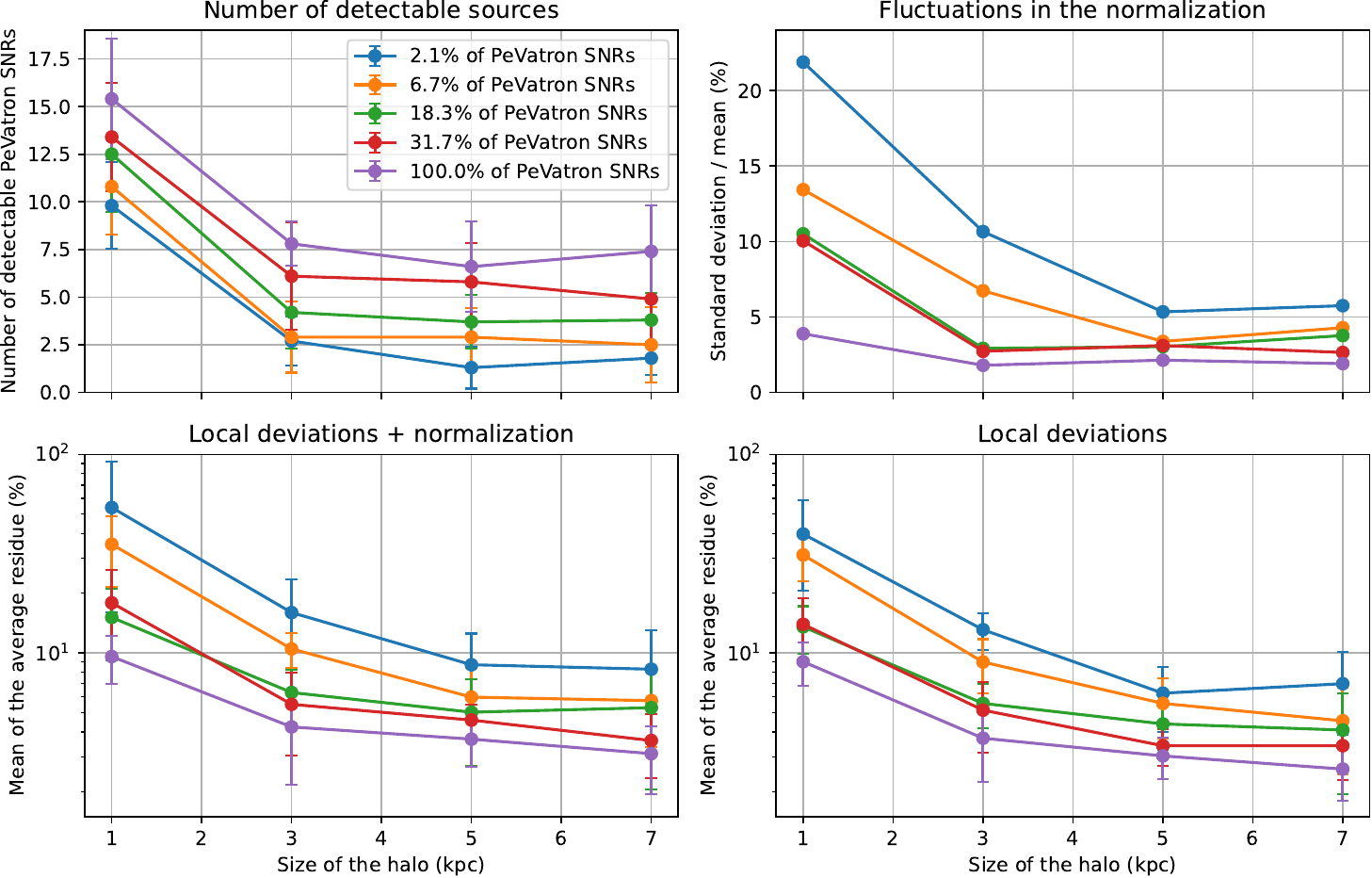}
    \caption{Dependence of the number of detectable sources (upper left panel) and the level of fluctuations of the diffuse background (upper right and lower panels) on the size of the halo for different percentages of PeVatron SNRs assuming a standard diffusion coefficient  defined by eq. (\ref{effective_diffusion}).}
    \label{fig:halo}
\end{figure}

The size of the halo, which is degenerated with the value of the diffusion coefficient, is a key parameter of our model, as it can be directly related to the residence time of cosmic rays in the Galaxy. Therefore, for the sake of completeness we investigate the dependence of our results on the size of the halo (figure \ref{fig:halo}) which shows our results on the detectability of sources (figure \ref{fig:detected_sources}) and on the level of fluctuations of the diffuse background (figure \ref{fig:fluctuations}) as a function of the halo size.

The upper left panel displays the number of detectable sources as a function of the halo size for different percentages of PeVatron SNRs in the simulation and for the standard diffusion coefficient defined by eq. (\ref{effective_diffusion}). First, we observe that the number of detectable sources tends to decrease when the halo size increases, although this tendency becomes less marked for halo sizes $\gtrsim3\,\rm{kpc}$. This result is due to the linear dependence of the diffusion coefficient on the halo size. Indeed, assuming a halo size of $1\,\rm{kpc}$ instead of $5\,\rm{kpc}$ leads to a much slower diffusion of cosmic rays that allows to transition to a sub-luminal diffusion speed much earlier compared to the case with a larger halo size, resulting in a significant slow-down in the source growth, allowing it to remain visible for a longer time. At the same time, varying the halo size from $5\,\rm{kpc}$ to $3\,\rm{kpc}$ or $7\,\rm{kpc}$ leaves the diffusion coefficient roughly unchanged, which explains the very small decrease in the number of detectable sources for $H\gtrsim3\,\rm{kpc}$. Another way to view this result is through the simultaneous linear dependence of the diffusion coefficient on the size of the halo $H$ and on the cosmic-ray energy $E^{1/3}$ for a Kolmogorov turbulence of the Galactic magnetic field. In this context $2\,\rm{PV}$ cosmic rays diffusing with a halo size of $1\,\rm{kpc}$ are equivalent to $\sim10\,\rm{TV}$ cosmic rays diffusing with a halo size of $5\,\rm{kpc}$. This point of view also demonstrates the self-consistency of our approach, as our model predicts many more visible sources in this case, in agreement with observations in the $\rm{TeV}$ energy range. Nevertheless, the most important point to be noted here is that, for the standard homogeneous diffusion mechanism, varying the halo size does not alleviate the degeneracy between different cases assuming different percentages of SNRs that are PeVatrons in the simulation, as for each halo size the different curves and their error bars overlap each other. Concerning the inhomogeneous diffusion case, represented by eq. (\ref{timeDependant}), we argue that there are already so many detectable sources, that the fluctuations in their number owing to a different halo size, would not alter the discussion made in section IV.

The upper right and the lower panels display the different quantifiers of the level of fluctuations that have been introduced in section V. Here again, we observe a tendency of the stochastic fluctuations of the diffuse gamma-ray background to decrease when the size of the halo increases. This can also be explained through the residence time of cosmic rays in the Galaxy, that is $\tau\sim H^2/D\propto H$. In this context, the smaller the halo size is, the smaller the residence time will be. This has the consequence to give an even more clumpy sea of cosmic rays, because they remain in the Galaxy for shorter times after having been released by their sources. This in turn makes the density of cosmic rays in each location of the Galaxy strongly dependent on its recent past supernova activity, which induces large spatial fluctuations further enhancing the stochastic effects. This fact can also be viewed from a geometrical perspective. Indeed, if the cosmic rays diffuse isotropically around their sources, and the geometry of the Galaxy is cylindrical with the height of the cylinder being much smaller than its radius (which is the case in our model), particles will typically not be able to fill the Galaxy radially on scales larger than their propagation scale in the $z$ direction before experiencing strong escape from the Galaxy. This effect will be further amplified as the halo size gets smaller. This behavior is particularly interesting and might help to constrain the size of the halo, as higher levels of fluctuations would hint towards a smaller halo size. Finally, we expect this discussion to also hold for the case of inhomogeneous diffusion around the sources, as the diffuse background is mostly related to the sea of cosmic rays and to old and very extended sources.

\end{document}